\documentclass[useAMS,usenatbib]{mn2e}
\usepackage[dvips]{graphicx}
\usepackage{amsmath}
\usepackage{amsfonts}
\usepackage{amssymb}

\title[The Darkest World]{
Detection of visible light from the darkest world}
\author[Kipping \& Spiegel]{David M. Kipping$^{1}$ \& David S. Spiegel$^{2}$\\
$^{1}$Center for Astrophysics, 60, Garden Street, Cambridge, MA 02138 [E-mail: 
dkipping@cfa.harvard.edu] \\
$^{2}$Department of Astrophysical Sciences, Peyton Hall, Princeton University, 
Princeton, NJ 08544}

\begin{document}

\date{Accepted 2011 August 1. Received 2011 July 25; in original form 2011 June 29}

\pagerange{\pageref{firstpage}--\pageref{lastpage}} \pubyear{2011}

\maketitle

\label{firstpage}

\begin{abstract}
We present the detection of visible light from the planet TrES-2b, the 
darkest exoplanet currently known. By analysis of the orbital 
photometry from publicly available \emph{Kepler} data (0.4-0.9\,$\mu$m), we 
determine a day-night contrast amplitude of $6.5\pm1.9$\,ppm, constituting 
the lowest amplitude orbital phase variation discovered.
The signal is detected to 3.7$\sigma$ confidence and persists in six
different methods of modelling the data and thus appears robust. In
contrast, we are unable to detect ellipsoidal variations 
or beaming effects, but we do provide confidence 
intervals for these terms.  If the day-night contrast is interpreted 
as being due to scattering, it corresponds to a geometric albedo of 
$A_g = 0.0253 \pm 0.0072$. However, our models indicate that there is 
a significant emission component to day-side brightness, and the true 
albedo is even lower ($<1$\%). By combining our measurement with 
\emph{Spitzer} and ground-based data, we show that a model with 
moderate redistribution ($P_n \simeq 0.3$) and moderate extra optical 
opacity ($\kappa^{\prime} \simeq0.3-0.4$) provide a compatible 
explanation to the data.
\end{abstract}


\vspace{-6mm}
\section{Introduction}
\label{sec:intro}

Orbital photometric phase variations have long been used in the study
and characterisation of eclipsing binaries \citep{wilson:1994}, where
the large masses and small orbital radii result in phase variations at
the magnitude to millimagnitude level. The three dominant components
of these variations are i) ellipsoidal variations, due to the
non-spherical nature of a star caused by gravitational distortion 
(e.g. \citeauthor{welsh:2010} \citeyear{welsh:2010}) ii)
relativistic beaming, due to the radial motion 
of the star shifting the stellar spectrum (e.g. \citealt{maxted:2000}) 
iii) reflected/emitted light, which varies depending on what phase of a
body is visible (e.g. \citeauthor{for:2010} \citeyear{for:2010}).

The visible bandpass orbital phase variations of a star due to a
hot-Jupiter companion are much smaller - around the part-per-million
(ppm) level - and thus have eluded detection until relatively
recently. The high precision space-based photometry of CoRoT 
(0.56-0.71\,$\mu$m) \citep{baglin:2009} and \emph{Kepler} 
\citep{basri:2005} have opened up this exciting new way of 
studying exoplanets for first time, with several detections recently 
reported in the literature:

\begin{itemize}
\item[{\tiny$\blacksquare$}] CoRoT-1b \citep{snellen:2009}; 
reflected/emitted light amplitude $126\pm36$\,ppm 
%
\item[{\tiny$\blacksquare$}] HAT-P-7b \citep{welsh:2010}; 
ellipsoidal amplitude $37$\,ppm, 
reflected/emitted light amplitude $63.7$\,ppm 
\item[{\tiny$\blacksquare$}] CoRoT-3b \citep{mazeh:2010}; 
ellipsoidal amplitude $(59\pm9)$\,ppm, 
beaming amplitude $(27\pm9)$\,ppm 
%
\item[{\tiny$\blacksquare$}] Kepler-7b \citep{demory:2011}; 
reflected/emitted light amplitude $(42\pm4)$\,ppm 
\end{itemize}

In this letter, we investigate the hot-Jupiter 
orbiting the G0V star TrES-2 \citep{odonovan:2006}, where we detect 
a reflected/emitted light amplitude of $(6.5 \pm 1.9)$\,ppm to a 
confidence of $3.7\sigma$, or 99.98\%. We also
measure the ellipsoidal variation and relativistic beaming amplitudes
to be $(1.5 \pm 0.9)$\,ppm and $(0.2\pm 0.9)$\,ppm respectively, 
which are broadly consistent with theoretical expectation.

If our detected signal is interpreted as being purely due to 
scattering, then the corresponding geometric albedo would be
$A_g = 0.0253 \pm 0.0072$ (using system parameters from Table 2, column 2 of
\citealt{kippingbakos:2011b} (KB11), as will be done throughout this work), 
meaning that just four months of \emph{Kepler}'s exquisite photometry 
has detected light from the darkest exoplanet yet found. Extrapolating 
to a 6\,year baseline, one can expect to detect albedos $\geq0.1$ 
(to 3$\sigma$ confidence) at similar orbital radii down to 
$R_P \simeq 3.0$\,$R_{\oplus}$. This clearly highlights the 
extraordinary potential which would be granted by an extended 
mission for \emph{Kepler}.

\vspace{-6mm}
\section{Observations \& Analysis}
\label{sec:observations}

\subsection{Data Acquisition}
\label{sub:dataacquisition}

We make use of ``Data Release 3'' (DR3) from the \emph{Kepler
Mission}, which consists of quarters 0, 1 and 2 (Q0, Q1 \& Q2). Full
details on the data processing pipeline can be found in the DR3
handbook. The data includes the use of BJD (Barycentric Julian Date)
time stamps for each flux measurement, which is crucial for time
sensitive measurements. All data used are publicly available via MAST.

We use the ``raw'' (labelled as ``AP\_RAW\_FLUX'' in the header) 
short-cadence (SC) data processed by the DR3 pipeline and a detailed 
description can be found in the accompanying release notes. The 
``raw'' data has been subject to PA (Photometric Analysis), which 
includes cleaning of cosmic ray hits, Argabrightenings, removal of 
background flux, aperture photometry and computation of centroid 
positions. It does not include PDC (Pre-search Data Conditioning) 
algorithm developed by the DAWG (Data Analysis Working Group). As 
detailed in DR3, this data is not recommended for scientific use, 
owing to, in part, the potential for under/over-fitting of the 
systematic effects.

\vspace{-4mm}
\subsection{Cleaning of the Data}
\label{sub:cleaning}

The raw data exhibit numerous systematic artifacts, including pointing
tweaks (jumps in the photometry), safe mode recoveries (exponential
decays) and focus drifts (long-term trends). The first effect may be
corrected by applying an offset surrounding the jump, computed using a
30-point interpolative function either side. Due to the exponential
nature of the second effect, we chose to exclude the affected data
rather than attempt to correct it. The third effect may be corrected
for using a detrending technique.

\begin{figure}
\begin{center}
\includegraphics[width=8.4 cm]{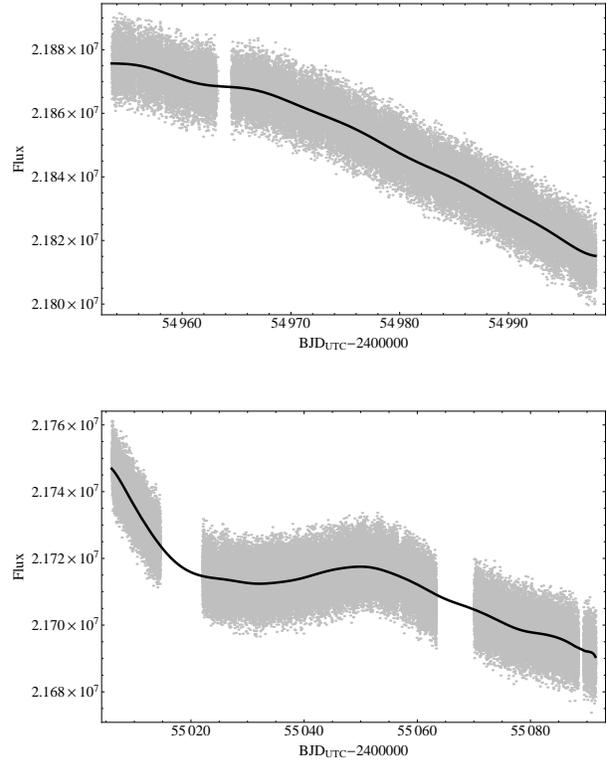}
\caption{\emph{``Raw'' (PA output) flux from DR3 of the \emph{Kepler}
pipeline for Q0\&Q1 (upper panel) and Q2 (lower panel) of the star
TrES-2. Overlaid is our model for the long-term trend, computed using
a discrete cosine transform for each data set. Outliers and
discontinuous systematic effects have been excluded.}}
\label{fig:detrending}
\end{center}
\end{figure}

For this latter effect, we use the cosine filter utilised to detect
ellipsoidal variations in CoRoT data by \citet{mazeh:2010}. The
technique acts as a high-pass filter allowing any frequencies of the
orbital period or higher through and all other long-term trends are
removed. Thus, we protect any physical flux variations on the time
scale of interest. We applied the filter independently to Q0+Q1 data
and then Q2 data. This is because the \emph{Kepler} spacecraft was
rotated in the intervening time and so the long-term trend will not be
continuous over this boundary. After removing 3$\sigma$ outliers with 
a running 20-point median and transits using the ephemeris of KB11, 
we applied the filter, with the resulting fitted trends shown in 
Fig~\ref{fig:detrending}. Our final cleaned data consists of 154,832 
SC measurements with a mean SNR$\simeq4408$.

\vspace{-4mm}
\subsection{Three Models}
\label{sub:3models}

We first define our null hypothesis, $\mathcal{M}_1$, where we
employ a flat line model across the entire time series,
described by a constant $a_0$.
For a physical description of the orbital phase variations, we first
tried the same model as that used by \citet{sirko:2003} and
\citet{mazeh:2010}. This simple model is sufficient for cases where
one is dealing with low signal-to-noise and reproduces the broad
physical features. The model, $\mathcal{M}_2$, is given by

\begin{align}
\mathcal{M}_2(\phi) &= a_0 + 0.5 a_{1c} \cos(\phi) + a_{1s} \sin(\phi) \nonumber \\
\qquad& + a_{2c} \cos(2\phi) + a_{2s} \sin(2\phi) \, ,
\label{eqn:model}
\end{align}

where $\phi$ is the orbital phase (defined as being 0 at the time of
transit minimum) and $a_{i}$ are coefficients related to the physical
model. $a_0$ is simply a constant to remove any DC (direct-current) 
component in the data. $a_{1c}$ corresponds to the reflection/emission 
effect and is expected to be have a negative amplitude. $a_{1s}$ 
corresponds to the relativistic beaming effect and is expected to be 
positive. $a_{2c}$ corresponds to the ellipsoidal variations and should be
negative. $a_{2s}$ is a dummy term which should be zero and ensures
the ellipsoidal variation is detected with the correct phase.

We also tried a third model, $\mathcal{M}_3$, where the
$a_{1c}$ term is replaced by the reflection caused by a
Lambertian sphere:

\begin{align}
0.5 a_{1c} \cos \phi \rightarrow a_{1c} \Bigg[ \frac{\sin|\phi|+(\pi-|\phi|)\cos|\phi|}{\pi} \Bigg].
\end{align}

%

\vspace{-4mm}
\subsection{Three Data Modes}
\label{sub:3data}

In addition to three models, we have three data input modes. The first
is simply corrected for detrending and nothing else, denoted
$\mathcal{D}_1$.  The second mode renormalises each orbital period
epoch. This renormalisation is done by computing the median of each
epoch and dividing each segmented time series by this value and we
denote this mode as $\mathcal{D}_2$. Finally, we tried allowing each
orbital period epoch to have its own variable renormalisation
parameter, which is simultaneously fitted to the data along with the
orbital phase curve model. This parameter is dubbed $a_{0,j}$ for the
$j^{\mathrm{th}}$ orbital period epoch. Denoting this data input mode
as $\mathcal{D}_3$, the fits now include an additional 51 free
parameters.


The models are fitted to the unbinned data using a Markov Chain 
Monte Carlo algorithm described in KB11 (method A) with
$1.25\times10^5$ accepted trials burning out the first 25,000. In
total, there are nine ways of combining the three models with the
three data modes. All nine models are fitted and results are given in
Table~\ref{tab:results}, with our preferred model description being
$\mathcal{M}_2$, $\mathcal{D}_3$ (since thermal emission is likely 
dominant over scattering, see \S\ref{sec:discussion}).

\vspace{-6mm}
\section{Results}
\label{sec:results}

\subsection{Orbital Photometry}
\label{sub:orbital}

Table~\ref{tab:results} presents the results of fitting the detrended
\emph{Kepler} photometry. Our models make no prior
assumption on the sign or magnitude of the $a_{i}$ coefficients. The
orbital period and transit epoch are treated as Gaussian priors from
the circular orbit results of KB11. 

\begin{table*}
\caption{\emph{Results of three models with three data modes, giving
nine sets of results. Emboldened row denotes our favoured solution.
Results do not include the orbital period, $P$
and transit epoch $\tau$, which are treated as Gaussian priors via $P
= 2.47061896 \pm 0.00000022$\,days and $\tau =
2454950.822014\pm0.000027$\,$\mathrm{BJD}_{\mathrm{TDB}}$.  Quoted
values are medians of marginalised posteriors with errors given by 
1$\sigma$ quantiles. $^{*}$ = parameter was fixed.  We do not show 
the $a_{0,j}$ fitted terms, which are simply renormalisation constants 
and are available upon request.}} 
\centering 
\begin{tabular}{c c c c c c} 
\hline\hline 
Model $\mathcal{M}$,  & $a_{1c}$ [ppm] & $a_{1s}$ [ppm] & $a_{2c}$ [ppm] & $a_{2s}$ [ppm] & $\chi^2$ \\ [0.5ex] 
Data $\mathcal{D}$ & (reflec./emiss.) & (beaming) & (ellipsoidal) & (dummy) & \\
\hline 
$\mathcal{M}_1$, $\mathcal{D}_1$ & $0^{*}$ & $0^{*}$ & $0^{*}$ & $0^{*}$ & 162603.5431 \\
$\mathcal{M}_2$, $\mathcal{D}_1$ & $-7.2_{-1.8}^{+1.8}$ & $0.78_{-0.85}^{+0.85}$ & $-1.42_{-0.92}^{+0.91}$ & $-0.27_{-0.85}^{+0.85}$ & 162583.4014 \\
$\mathcal{M}_3$, $\mathcal{D}_1$ & $-7.3_{-1.9}^{+1.8}$ & $0.79_{-0.86}^{+0.86}$ & $-0.77_{-0.91}^{+0.92}$ & $-0.26_{-0.86}^{+0.86}$ & 162583.3162\\
\hline
$\mathcal{M}_1$, $\mathcal{D}_2$ & $0^{*}$ & $0^{*}$ & $0^{*}$ & $0^{*}$ & 161875.4005 \\
$\mathcal{M}_2$, $\mathcal{D}_2$ & $-6.4_{-1.8}^{+1.8}$ & $0.34_{-0.87}^{+0.86}$ & $-1.52_{-0.94}^{+0.93}$ & $0.19_{-0.87}^{+0.87}$ & 161859.6732 \\
$\mathcal{M}_3$, $\mathcal{D}_2$ & $-6.4_{-1.9}^{+1.8}$ & $0.34_{-0.86}^{+0.86}$ & $-0.95_{-0.92}^{+0.92}$ & $0.19_{-0.87}^{+0.86}$ & 161859.6095 \\
\hline
$\mathcal{M}_1$, $\mathcal{D}_3$ & $0^{*}$ & $0^{*}$ & $0^{*}$ & $0^{*}$ & 161837.6648 \\
$\mathbf{\mathcal{M}_2}$, $\mathbf{\mathcal{D}_3}$ & $\mathbf{-6.5_{-1.9}^{+1.9}}$ & $\mathbf{0.22_{-0.87}^{+0.88}}$ & $\mathbf{-1.50_{-0.93}^{+0.92}}$ & $\mathbf{0.31_{-0.87}^{+0.88}}$ & $\mathbf{161821.7228}$ \\
$\mathcal{M}_3$, $\mathcal{D}_3$ & $-6.7_{-1.8}^{+1.8}$ & $0.23_{-0.88}^{+0.89}$ & $-0.90_{-0.91}^{+0.91}$ & $0.32_{-0.88}^{+0.88}$ & 161821.7232 \\
\hline
Theory Expectation & $-20\rightarrow0$ & $\sim2.4$ & $\sim-2.3$ & 0 & - \\
\hline\hline 
\end{tabular}
\label{tab:results} 
\end{table*}

\begin{figure*}
\begin{center}
\includegraphics[width=16.8 cm]{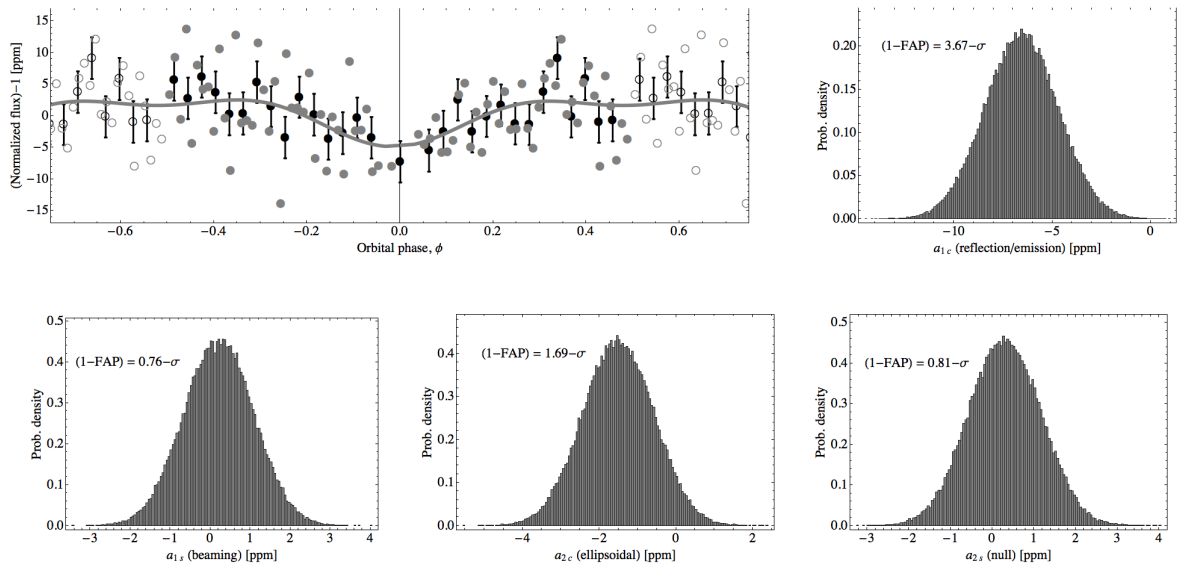}
\caption{\emph{
\textbf{Top Left}: Final fit to the phased photometry. Points without
errors are the 2000-point phase binned data. Points with errors are
5000-point phase binned data. Best-fit model $\mathcal{M}_2$ with data mode
$\mathcal{D}_3$ shown in solid. Note that all fits were performed on the
unbinned photometry.
\textbf{Top Right \& Lower Panels}: Marginalised posterior 
distributions for the same model of four fitted parameters. Unity 
minus the false-alarm-probability values are provided for each 
parameter, based upon an odds ratio test described in 
\S\ref{sub:orbital}.}}
\label{fig:super}
\end{center}
\end{figure*}

When considering statistical significance, what one is really
interested in is the confidence of detecting each physical effect
i.e. reflection/emission, ellipsoidal and beaming. For this reason,
model comparison tools, such as the Bayesian Information Criterion
(BIC) or an F-test are inappropriate. This is because these methods
evaluate the preference of one hypothesis over another, where the two
models would be a null-hypothesis and a hypothesis including
reflection/emission, ellipsoidal variations, beaming and the dummy
term. Thus, any inference drawn from this would be for the
\emph{entire} model and not for each individual effect. In the
analysis presented here, simple inspection of the posteriors from
Fig~\ref{fig:super} shows that only one effect is actually detected 
(reflection/emission), but a model comparison method would evaluate 
the significance of all four physical effects (including the dummy 
term) versus no effect.

A more useful statistical test would consider the significance of each
physical effect individually from a joint fit. An excellent tool to
this end is the odds ratio test discussed in \citet{hatp24:2010}.  If
a parameter was equal to zero, we would expect 50\% of the MCMC runs
to give a positive value and 50\% to give a negative value.  Consider
that some asymmetry exists and a fraction $f$ of all MCMC trial were
positive and $1-f$ were negative. The reverse could also be true and
so we define $f$ such that $f>0.5$ i.e. it represents the majority of
the MCMC trials. The odds ratio of the asymmetric model over the 50:50
model is:

\begin{equation}
\mathrm{O} = \frac{0.5}{1-f}
\end{equation}

For only two possible models, the probability of the asymmetric model
being the correct one is $\mathrm{P}(\mathrm{asym}) = 1 - [1/(1 +
\mathrm{O})]$. We perform this test on each of the four parameters
fitted for, $a_{1c}$, $a_{2c}$, $a_{1s}$ and $a_{2s}$. The associated
results are visible in the top-left corners of each posterior shown in
Fig~\ref{fig:super}, for our preferred model and data mode
i.e. $\mathcal{M}_2$, $\mathcal{D}_3$. To summarise, only one
parameter presents a significant detection - the reflection/emission
effect. Here, we find $a_{1c}$'s posterior is sufficiently asymmetric
to have a probability of occurring by random chance of just 0.02\%,
which equates to 3.67$\sigma$. We consider any signal detected above
3$\sigma$ confidence to merit the claim of a ``detection'' rather
than a measurement and thus we find TrES-2b to be the darkest
exoplanet from which visible light has been detected.

As discussed in \S\ref{sub:3models}, we tried two different models for
the reflection/emission effect; a simple sinusoid ($\mathcal{M}_2$)
and the reflected light from a perfectly Lambertian sphere
($\mathcal{M}_3$). Between the two models, there is negligible
difference in the goodness-of-fit, as seen in Table~\ref{tab:results},
for all three data modes. Including the Lambertian model
takes some power away from the ellipsoidal variations though and thus
the current data does not yield a preference 
between a Lambertian sphere model or stronger ellipsoidal variations.

\vspace{-4mm}
\subsection{Occultation Measurement}
\label{sub:occultation}

The duration of the transit, and thus occultation since TrES-2b
maintains negligible eccentricity, is equal to
$4624\pm42$\,seconds (defined as the time
between when the planet's centre crosses the stellar limb to exiting
under the same condition). In contrast, the orbital period of TrES-2b
is $2.470619$\,days. We therefore obtain $\sim$46 times more
integration time of the orbit than the occultation event. This
indicates that we should expect to be able to reach a sensitivity of 
$\sqrt{46}$ times greater, purely from photon statistics. The 
uncertainty on our phase curve measurement is $\pm1.9$\,ppm.  We 
therefore estimate that one should have an uncertainty on the 
occultation depth of $\sim13$\,ppm. If we assume the nightside has a 
negligible flux, then the depth of the occultation is expected to be 
$6.5$\,ppm (i.e. equal to the day-night contrast), and this already 
suggests that the present publicly available \emph{Kepler} photometry 
will be insufficient to detect the occultation.
To test this hypothesis, we will here fit the occultation
event including the Q0, Q1 and Q2 data.

To perform our fit, we use the same Gaussian priors on $P$ and
$\tau$ as earlier. We also adopt priors for other important system 
parameters from KB11, such as 
$b=0.8408\pm0.0050$, $p^2 = 1.643\pm0.067$\% and $\widetilde{T}_1 = 
4624\pm42$\,seconds. We stress that these are all priors and not 
simply fixed parameters. We also make use of the priors on the 
$a_{0,j}$ coefficients from the $\mathcal{M}_3$, $\mathcal{D}_3$ fit. 
Data are trimmed to be within $\pm0.06$\,days of the expected time of 
occultation to prevent the phase curve polluting our signal, leaving 
us with 8457 SC measurements. Assuming a circular orbit, the data were 
fitted using an MCMC algorithm.

The marginalised posterior of the
occultation depth yields $\delta_{\mathrm{occ}} =
16_{-14}^{+13}$\,ppm, which is clearly not a significant detection.
The derived uncertainty of 13-14\,ppm is very close to our estimation 
of $\sim13$\,ppm and thus supports our hypothesis that the current
\emph{Kepler} data are insufficient to detect the occultation of
TrES-2b.  We also note that the inclusion of the Q2 data does improve
the constraints on the occultation event (KB11 found 
$\delta_{\mathrm{occ}} = 21_{-22}^{+23}$\,ppm using Q0 \& Q1 only).

\vspace{-6mm}
\section{Discussion}
\label{sec:discussion}

Hot-Jupiters are generally expected to be dark.  Significant
absorption due to the broad wings of the sodium and potassium D lines
is thought to dominate their visible spectra \citep{sudarsky:2000},
leading to low albedos of a few percent.  Indeed, aside from
the recent report of Kepler-7b's $(38\pm12)$\% \emph{Kepler}-band 
geometric albedo (KB11), searches for visible 
light from hot-Jupiters have generally revealed mere upper limits 
(\citealt{collier:2002,leigh:2003,rowe:2008,burrows:2008}). 

The 6.5$\pm$1.9\,ppm contrast (determined from our preferred 
model $\mathcal{M}_2$, $\mathcal{D}_3$) between the day-side and night-side
photon flux that we measure for TrES-2b represents the
most sensitive measurement yet of emergent radiation in the visible
from a hot-Jupiter, and is a factor of $\sim$20 and $\sim$6 dimmer
than the corresponding differences for HAT-P-7b \citep{welsh:2010}
and Kepler-7b.

In order to interpret the visible flux, we use the planetary
atmosphere modelling code {\tt COOLTLUSTY} \citep{hubeny:2003}.  For
simplicity, we adopt equilibrium chemistry with nearly Solar abundance
of elements, although we leave titanium oxide and vanadium oxide (TiO
and VO) out of the atmosphere model.  These compounds could, if
present in the upper atmosphere of a hot-Jupiter, strongly affect the
atmosphere structure and the visible and near infrared spectra, by
making the atmosphere more opaque in the visible and by leading to a
thermal inversion if the stellar irradiation exceeds
$\sim$10$^9$$\rm~erg~cm^{-2}~s^{-1}$ \citep{hubeny:2003,fortney:2008}.
We leave TiO and VO out of our calculations, however, because of the
difficulty of maintaining heavy, condensible species high in the
atmospheres \citep{spiegel:2009}.  Instead, we use an ad hoc extra
opacity source $\kappa^{\prime}$, as described in \citet{spiegel:2010}.

We calculate a grid of models with $\kappa^{\prime}$ ranging (in
$\rm~cm^2~g^{-1}$) from 0 to 0.6 in steps of 0.1 and with
redistribution $P_n$ ranging from 0 to 0.5 in steps of 0.1 ($P_n$
represents the fraction of incident irradiation that is transported to
the nightside, which is assumed in our models to occur in a pressure
range from 10 to 100\,mbars).  For each of these 42 parameter
combinations, we calculate a day-side model, a night-side model, and a
model that has the same temperature/pressure structure as the dayside
but that has the star turned off, so as to calculate the emitted (and
not scattered) flux (thus also giving the scattered component).

We draw several inferences from our models and the data.
First, the nightside contributes negligible flux in the 
\emph{Kepler}-band (always $<$12\% of the dayside, and for most models 
significantly less than that), meaning that the 6.5\,ppm number represents 
essentially the entire day-side flux.

\begin{figure}
\begin{center}
\includegraphics[width=8.4 cm]{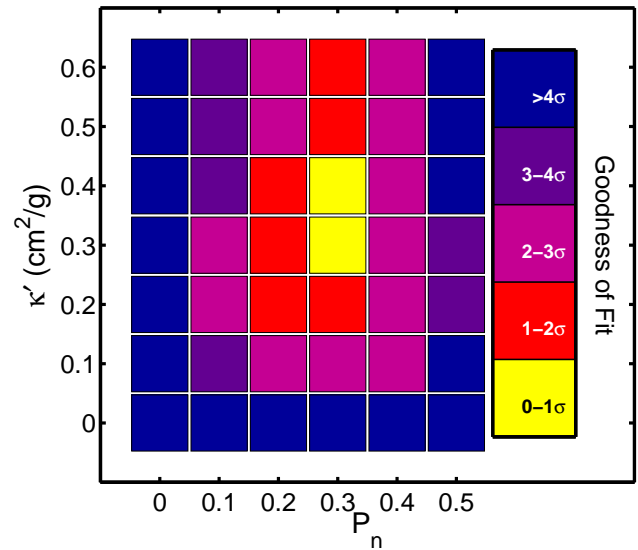}
\caption{\emph{Goodness-of-fit for a grid of atmosphere models. The
models that are consistent with available \emph{Kepler}-band, 
\emph{Ks}-band, and \emph{Spitzer} IRAC data have moderate 
redistribution to the night side ($P_n$) and moderate extra optical 
opacity ($\kappa^{\prime}$). Models with $\kappa^{\prime}=0$ can be 
ruled out on the basis of the \emph{Kepler} data alone.}}
\label{fig:Pnkappa}
\end{center}
\end{figure}

Second, by also including the available infrared secondary eclipse
data on TrES-2b \citep{odonovan:2010,croll:2010}, we find that in our
model set there must be some redistribution (but not too much) and
there must be some extra absorber (but not too much).  For each model,
we compute a $\chi^2$ value, including 6 data points: 
\emph{Kepler}-band, \emph{Ks}-band, and the four \emph{Spitzer} IRAC 
channels (3.6, 4.5, 5.8 \& 8.0$\mu$m).   
Fig~\ref{fig:Pnkappa} portrays the $\chi^2$ values of our grid of 
models, with the colour ranges corresponding to the $\chi^2$ values 
bounding 68.3\% of the integrated probability (1$\sigma$), 95.5\% 
(2$\sigma$), 99.7\% (3$\sigma$) and 99.99\% (4$\sigma$).  The models 
that best explain the available data correspond to $\kappa^{\prime}
\sim 0.3-0.4 \rm~cm^2~g^{-1}$ and $P_n\sim0.3$ ($\sim$30\% of incident 
flux redistributed to the night).  In particular, models with no extra
absorber are completely inconsistent with observations, even on the
basis of the \emph{Kepler} data alone.  The upshot is that some extra 
opacity source appears to be required to explain the emergent 
radiation from this extremely dark world.
Owing to this optical opacity, our models that are consistent with the
data have thermal inversions in their upper atmosphere, as in
\citet{spiegel:2010}.
We note that \citet{madhu:2010} find that the IR data of TrES-2b may be
explained by models both with and without thermal
inversions; nevertheless, we believe that optical opacity sufficient
to explain the \emph{Kepler} data is likely to heat the upper
atmosphere, as per \citet{hubeny:2003}.

Finally, by computing the scattered contribution to the total flux, we
find that for all parameter combinations the scattered light
contributes $\lesssim$10\% of the \emph{Kepler}-band flux, and for the
best-fit models the scattered light is $\lesssim$1.5\% of the total.
TrES-2b, therefore, appears to have an extremely low geometric albedo
(for all models, the geometric albedo is $<1$\%, and for the
best-fit models it is $\sim$0.04\%).
  Exact values for the 
  amount of extra optical opacity, redistribution and the albedo 
  cannot be presently provided because inferences about them depend on 
  unknown quantities such as the wavelength dependence of the extra 
  opacity source and the altitude dependence of winds.

\vspace{-6mm}
\section*{Acknowledgments}

We thank the \emph{Kepler} Science Team, especially
the DAWG, for making the data used here available. 
Thanks to A. Burrows, M. Nikku \& the anonymous referee for helpful comments and 
I. Hubeny \& A. Burrows for the development and continued maintenance 
of {\tt COOLTLUSTY} and associated opacity database.
DMK is supported by Smithsonian Instit. Restricted Endowment Funds.


\label{lastpage}


\vspace{-6mm}
\begin{thebibliography}{99}

\bibitem[\protect\citeauthoryear{Baglin et al.}{2009}]{baglin:2009} Baglin, A. 
et al. 2009, Transiting Planets, Proc. IAU Symp., 253, 71

\bibitem[\protect\citeauthoryear{Basri et al.}{2005}]{basri:2005} Basri, G., 
Borucki, W. J. \& Koch, D. 2005, New Astronomy Rev., 49, 478

\bibitem[\protect\citeauthoryear{Burrows et al.}{2008}]{burrows:2008}
Burrows, A., Ibgui, L., \& Hubeny, I.\ 2008, ApJ, 682, 1277

\bibitem[\protect\citeauthoryear{Demory et al.}{2011}]{demory:2011}
Demory, B.-O. et al. 2011, ApJL, accepted

\bibitem[\protect\citeauthoryear{Collier Cameron et al.}{2002}]{collier:2002} 
Collier Cameron, A., Horne, K., Penny, A. \& Leigh, C. 2002, MNRAS, 330, 187

\bibitem[\protect\citeauthoryear{Croll et al.}{2010}]{croll:2010}
Croll, B., Albert, L., Lafreniere, D., Jayawardhana, R., \& Fortney,
J.~J.\ 2010, ApJ, 717, 1084

\bibitem[\protect\citeauthoryear{For et al.}{2010}]{for:2010} For,
B.-Q., et al.  2010, ApJ, 708, 253

\bibitem[\protect\citeauthoryear{Fortney et al.}{2008}]{fortney:2008}
Fortney, J.~J., Lodders, K., Marley, M.~S., \& Freedman, R.~S.\ 2008,
ApJ, 678, 1419

\bibitem[\protect\citeauthoryear{Hubeny et al.}{2003}]{hubeny:2003}
Hubeny, I., Burrows, A., \& Sudarsky, D.\ 2003, ApJ, 594, 1011

\bibitem[\protect\citeauthoryear{Kipping et al.}{2010}]{hatp24:2010}
Kipping, D. M. et al. 2010, ApJ, 725, 2017

\bibitem[\protect\citeauthoryear{Kipping \&
Bakos}{2011}]{kippingbakos:2011b} Kipping, D. M. \& Bakos,
G. A. 2011, ApJ, 733, 36 (KB11)

\bibitem[\protect\citeauthoryear{Leigh et al.}{2003}]{leigh:2003} 
Leigh, C., Collier Cameron, A., Horne, K., Penny, A. \& James, D. 2003, 
MNRAS, 344, 1271

\bibitem[\protect\citeauthoryear{Madhusudhan \&
Seager}{2010}]{madhu:2010} Madhusudhan, N., \& Seager, S.\ 2010,
ApJ, 725, 261

\bibitem[\protect\citeauthoryear{Maxted et al.}{2000}]{maxted:2000} 
Maxted, P. F. L., Marsh, T. R. \& North,
R. C. 2000, MNRAS, 317, L41

\bibitem[\protect\citeauthoryear{Mazeh \& Faigler}{2010}]{mazeh:2010}
Mazeh, T.  \& Faigler, S. 2010, A\&A, 521, 59

\bibitem[\protect\citeauthoryear{O'Donovan et
al.}{2006}]{odonovan:2006} O'Donovan, F.~T. et al. 2006, ApJ, 651, L61

\bibitem[\protect\citeauthoryear{O'Donovan et
al.}{2010}]{odonovan:2010} O'Donovan, F.~T., Charbonneau, D.,
Harrington, J., Madhusudhan, N., Seager, S., Deming, D. \& Knutson,
H.~A.\ 2010, ApJ, 710, 1551

\bibitem[\protect\citeauthoryear{Rowe et al.}{2008}]{rowe:2008}
Rowe, J.~F. et al.\ 2008, ApJ, 689, 1345

\bibitem[\protect\citeauthoryear{Sirko \&
Paczynski}{2003}]{sirko:2003} Sirko, E. \& Paczynski, B. 2003, ApJ,
592, 1217

\bibitem[\protect\citeauthoryear{Snellen et al.}{2009}]{snellen:2009}
Snellen, I. A. G., de Mooij, E. J. W. \& Albrecht, S. 2009, Nature, 459, 543

\bibitem[\protect\citeauthoryear{Spiegel et al.}{2009}]{spiegel:2009}
Spiegel, D.~S., Silverio, K. \& Burrows, A.\ 2009, ApJ, 699, 1487

\bibitem[\protect\citeauthoryear{Spiegel \&
Burrows}{2010}]{spiegel:2010} Spiegel, D. S. \& Burrows, A. 2010, ApJ,
722, 871

\bibitem[\protect\citeauthoryear{Sudarsky et
al.}{2000}]{sudarsky:2000} Sudarsky, D., Burrows, A., \& Pinto, P.
2000, ApJ, 538, 885

\bibitem[\protect\citeauthoryear{Welsh et al.}{2010}]{welsh:2010}
Welsh, W. F., Orosz, J. A., Seager, S., Fortney, J. J., Jenkins, J.,
Rowe, J. F., Koch, D. \& Borucki, W. J. 2010, ApJ, 713, 145

\bibitem[\protect\citeauthoryear{Wilson}{1994}]{wilson:1994} Wilson,
R. E. 1994, PASP, 106, 921

\end{thebibliography}
\end{document}